\documentclass[
preprint,
amsmath,amssymb,
aps,
prl,
nolongbibliography,
floatfix,
]{revtex4-2}

\usepackage{color}
\def\cbl{\color{black}}
\def\cb{\color{black}}
\usepackage{float}
\usepackage{graphicx}
\usepackage{dcolumn}
\usepackage{bm}
\usepackage{hyperref}
\usepackage[mathlines]{lineno}

\usepackage{setspace}

\begin{document}

\title{Dimension-dependent continuum limits in tissue mechanics}

\author{Matthew J. Simpson}
\email{matthew.simpson@qut.edu.au}
\author{Pascal R. Buenzli}
\affiliation{School of Mathematical Sciences, Queensland University of Technology (QUT), Brisbane, Australia.}

\date{\today}

\begin{abstract}
Continuum descriptions of epithelial tissue mechanics can replace expensive individual-based simulations with tractable macroscopic models, yet the link between cell-scale forces and tissue-scale transport remains poorly understood. We show that dimensionality controls this link: long-time mechanical relaxation rates reveal generalized porous-media-type nonlinear transport phenomena, $D(\rho)\propto\rho^\gamma$. Exponents in nonlinear diffusivities are fixed by microscopic mechanics and dimensionality, providing a novel physical mechanism for emergent macroscopic transport phenomena.
\end{abstract}

\maketitle
\newpage  
Discrete, cell-based models link microscopic forces to tissue-scale evolution~\cite{Odell1981,Oster1983,PittFrancis2009,Mirams2013}, but are computationally intractable with large cell numbers and difficult to parameterise. Continuum models offer a practical alternative, enabling density-level parameter inference while revealing emergent links between microscopic mechanisms and emergent tissue-scale transport~\cite{Simpson2025}. Although 1D continuum limits have been widely studied, the corresponding 2D problem remains largely unexplored~\cite{Fozard2009,Murray2009,Murray2011,Murray2012,Murphy2019,Murphy2020}.  As we will show, this matters because dimension fundamentally alters the macroscopic transport phenomena, as it does in other physical systems~\cite{Hughes1995,Redner2001,Goldenfeld1990,Bricmont1991}. In particular, epithelial mechanics motivates porous-medium-type nonlinear diffusivities, $D(\rho)\propto\rho^\gamma$~\cite{Vazquez2007,Buenzli2025}.  These models include both $\gamma > 0$ and $\gamma < 0$, unlike most phenomenological applications in fluid mechanics~\cite{Bear1972,Barenblatt1996} and population biology~\cite{OkuboLevin2001,Murray2002} that are limited to $\gamma > 0$.  We establish that $\gamma$ is determined jointly by the microscopic force law and spatial dimension. Our approach uses long-time exponential relaxation: by matching relaxation rates in discrete mechanical models to those of macroscopic partial differential equation (PDE) models near steady state, we infer novel continuum diffusion laws. The continuum models thereby coarse-grain the microscopic mechanics, identifying which force-law features persist macroscopically, and identifying how dimension reshapes them into an effective diffusion exponent.\\


\noindent
\textit{Discrete model.}  

\begin{figure}[htp]
  \centering
\includegraphics[width=1\columnwidth]{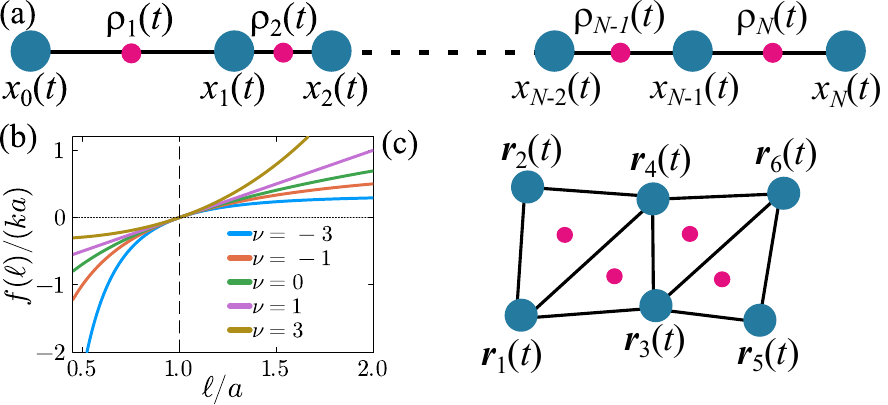}
\caption{\cbl Discrete models and force-length laws. (a) 1D chain with fixed endpoints, interval lengths $\ell_n(t)$, and densities $\rho_n(t)=1/\ell_n(t)$. (b) Force--length laws, $f(\ell)=k(\ell^\nu-a^\nu)/(\nu a^{\nu-1})$ for $\nu\neq0$, with logarithmic limit $f(\ell)=ka\log(\ell/a)$. (c) 2D network of moving vertices connected by pairwise mechanical interactions. \cb }
\label{fig:F1}
\end{figure}

In 1D the tissue is modelled as a chain of $N$ mechanical links representing cells (Figure \ref{fig:F1}a). We impose fixed boundaries, $x_0(t)=0$, $x_{N}(t)=L>0$, and evolve interior nodes according to an overdamped model
\begin{equation}\label{eq:mechanical-system}
\eta x'_n(t)  
= f(\ell_{n}) - f(\ell_{n-1}), 
\quad n=1,2,\ldots,N-1,
\end{equation}
where $\eta > 0$ is a drag coefficient, $\ell_n(t) = x_{n+1}(t) - x_{n}(t)>0$, and the prime denotes differentiation with respect to time $t$.  This model is closed by specifying an interaction force, $f(\ell)$.   Provided $f(\ell)$ is well-behaved (see below), time-dependent solutions of Equation \eqref{eq:mechanical-system} approach an equally-spaced steady state, $\bar{x}_n = nL/N$ for $n=0,1,2,\ldots,N$~\cite{Buenzli2025}.

Time-dependent solutions of Equation \eqref{eq:mechanical-system} depend on $f(\ell)$, with the simplest and most well-studied case being Hooke's law, $f(\ell) = k (\ell - a)$, where $a > 0$ is the resting cell length.  While Equation \eqref{eq:mechanical-system} can be solved exactly for Hooke's law~\cite{Buenzli2025}, we work more generally with a nonlinear system of ordinary differential equations (ODEs) corresponding to a parameterized family of interaction force laws 
\begin{equation}\label{eq:forcelaws}
f(\ell) =
\begin{cases}
k\left(\ell^\nu-a^\nu\right)/(\nu a^{\nu-1}),
& \nu \ne 0, \\[2ex]
ka\log\left(\ell/a\right),
& \nu = 0.
\end{cases}
\end{equation}
This choice has $f(a)=0$ and $f'(a) = k$ so that $f(\ell)$ linearizes to Hooke's law near $\ell=a$ for all $\nu$. For $k, a, \ell >0$, $f(\ell)$ is concave up for $\nu>1$, linear for $\nu=1$, and concave down for $\nu<1$ (Figure \ref{fig:F1}b).  Closed-form solutions of Equations \eqref{eq:mechanical-system}--\eqref{eq:forcelaws} are not tractable, however late time solutions can be obtained by linearizing about $\bar{\mathbf{x}}=(\bar{x}_0,\bar{x}_1,\ldots,\bar{x}_{N})^\top$.  Writing $\bar{\ell} = L/N$ and $\delta x_n(t)=x_n(t)-\bar{x}_n$ for 
 $n=0,1,\ldots,N$, we collect the interior perturbations as
$\boldsymbol{\delta x}(t)=(\delta x_1(t),\delta x_2(t),\ldots,\delta x_{N-1}(t))^\top$, noting that the interior nodes evolve linearly
\begin{equation} \label{eq:linearized}
\eta \delta x_n'(t)
=-\kappa\left(-\delta x_{n-1}+2\delta x_n-\delta x_{n+1}\right),
\end{equation}
for $n=1,2,\ldots,N-1$, $\delta x_0(t) = \delta x_N(t) = 0$, $\kappa =  f'(\bar{\ell}) =  k(\bar{\ell}/a)^{\nu-1}$. The solution of Equation \eqref{eq:linearized} is $\boldsymbol{\delta x}(t)  =  \sum_{p=1}^{N-1} a_p \exp{\left(-\kappa\lambda_p t / \eta  \right)} \mathbf{v}^{(p)}$, where $\mathbf{v}^{(p)}$ is the $p$th eigenvector of the $(N\!-\!1)\times (N\!-\!1)$ tridiagonal matrix associated with Equation \eqref{eq:linearized} with eigenvalues $\lambda_p=2-2\cos\left(\pi p/N\right)$ for $p=1,2,\ldots,N-1$~\cite{Kouachi2006}.  Since $\kappa/\eta>0$ and $\lambda_p>0$ for all $p$, $\bar{\mathbf{x}}$ is asymptotically stable.  The coefficients $a_p$ are determined by $\boldsymbol{\delta x}(0)$. For generic $\boldsymbol{\delta x}(0)$ with nonzero projection onto $\boldsymbol{v}^{(1)}$, the late-time relaxation to $\bar{\mathbf{x}}$ is governed by the leading eigenvalue, $\lambda_1 = 2-2\cos\left(\pi/N\right)$~\cite{Haberman2013,Simpson2015}. As $t \to \infty$, the magnitude of the perturbation decays like $\|\boldsymbol{\delta x}(t)\| \propto
\exp\left(-\kappa \lambda_1 t / \eta  \right)$ as $t\to\infty$.  In seeking a continuum limit description, it is natural to consider large $N$, noting that $\lambda_1 \sim (\pi/N)^2$ as $N \to \infty$ giving    
\begin{equation}\label{eq:DiscreteLateTime}
\|\boldsymbol{\delta x}(t)\| \propto
\exp\left[
-(k/\eta)
\left(\pi/N\right)^2
\left(\bar{\ell}/{a}\right)^{\nu-1}t
\right].
\end{equation}
Three factors control the relaxation time: $k/\eta$, $(\pi/N)^2$, and $(\bar{\ell}/a)^{\nu-1}$.  For the special case of Hooke's law ($\nu = 1$), the influence of $L$ and $a$ through $(\bar{\ell}/a)^{\nu-1}$ is hidden.  

\begin{figure}[!htbp]
  \centering
\includegraphics[width=0.65\columnwidth]{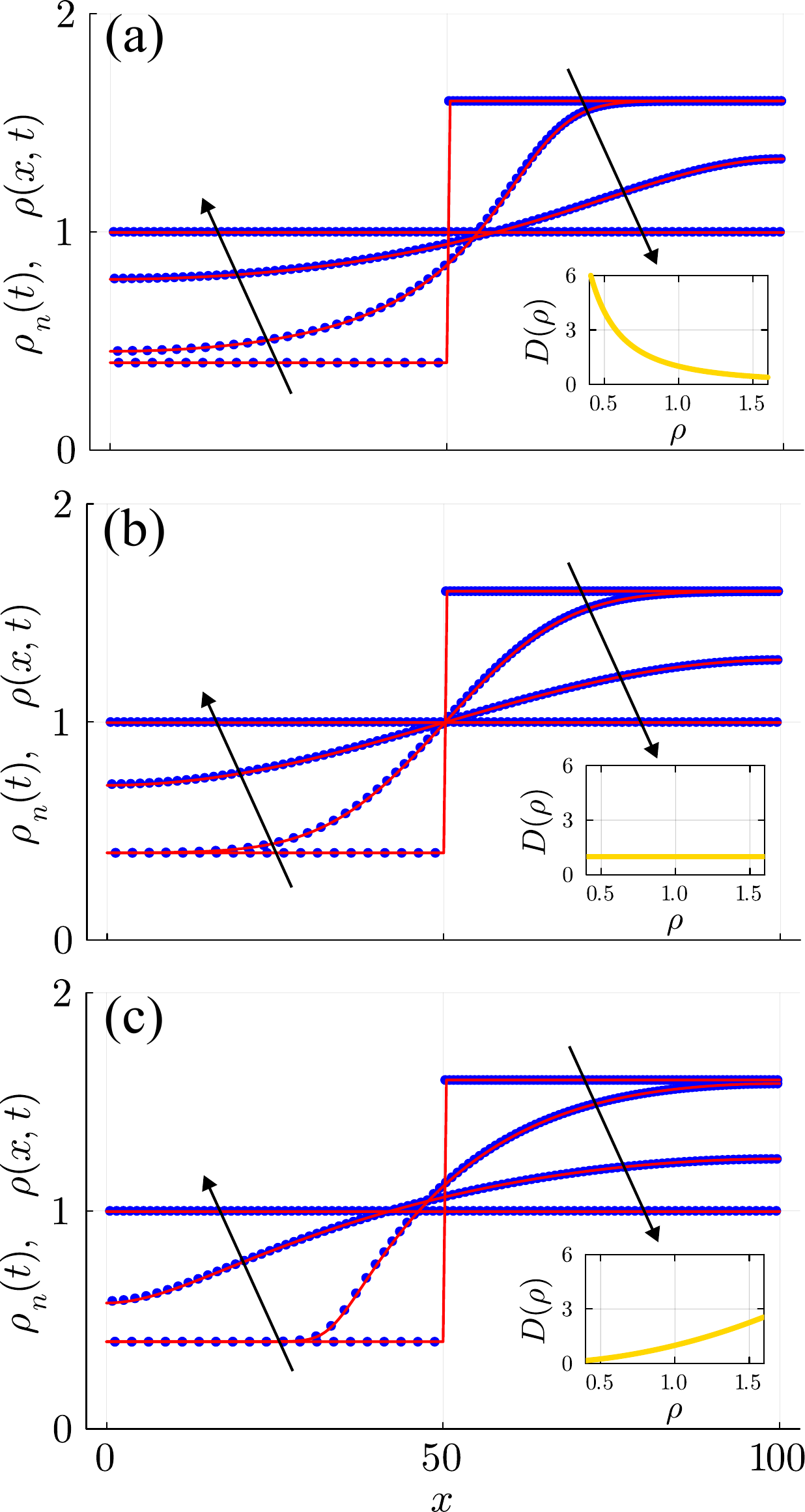}
\caption{\cbl 1D continuum--discrete comparisons. (a)--(c) Continuum densities $\rho(x,t)$ (solid red curves) are compared with discrete densities $\rho_n(t)=1/\ell_n(t)$ (blue dots) for the force law $f(\ell)=k(\ell^\nu-a^\nu)/(\nu a^{\nu-1}))$ with $\nu=1,-1,-3$, corresponding to $D(\rho)=\rho^{-2},\rho^0,\rho^2$, respectively. Profiles are shown at $t=0,10^2,10^3,10^4$, with arrows indicating increasing time, $t$.  Insets show $D(\rho)$. Parameters are $k=a=\eta=1$, $L=N=100$, with initial density $\rho=2/5$ for $x<50$ and $\rho=8/5$ for $x>50$. The mean density is $\bar{\rho}=N/L=1$, so the steady-state spring length satisfies $\bar{\ell}=1/\bar{\rho}=a=1$, corresponding to zero prestress at steady state. \cb}
\label{fig:F2}
\end{figure}

\begin{figure}[!htbp]
\centering
\includegraphics[width=0.60\columnwidth]{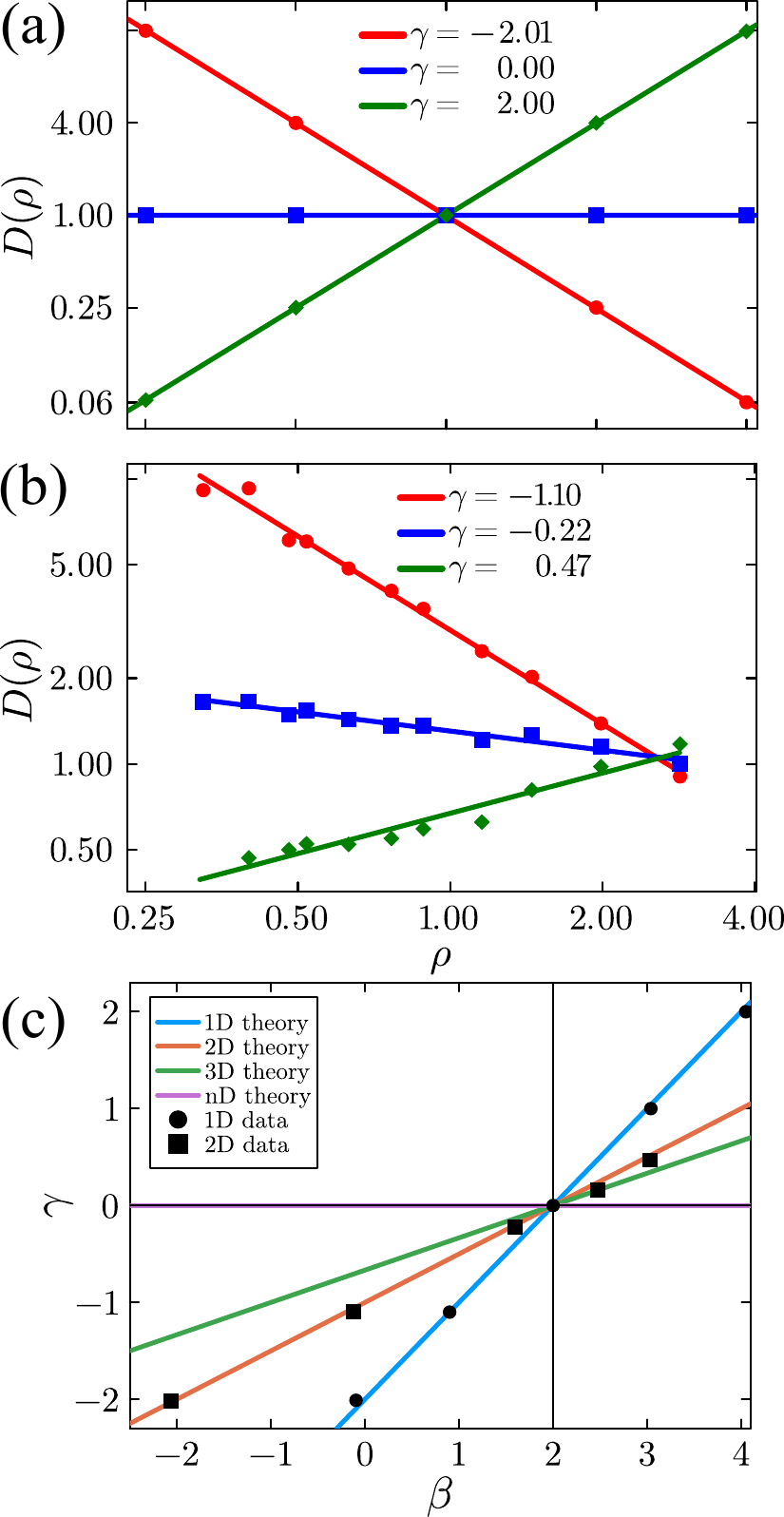}
\caption{\cbl Coarse-graining microscopic dynamics into $D(\rho)$ from relaxation rates. (a) In 1D, simulation-based estimates obtained over $0.5\leq\bar{\rho}\leq4$ give fitted exponents $\gamma=-2.01,0.00,2.00$ for $\nu=1,-1,-3$, respectively, consistent with Equation~\eqref{eq:1DDiffusivity}. Complementary simulations vary the resting length over $0.5\leq a\leq5$ for $\nu=1,-1$ and $0.5\leq a\leq4$ for $\nu=-3$. (b) Analogous 2D simulations give $\gamma=-1.10,-0.22,0.47$ for $\nu=1,-1,-3$, respectively, using $0.17\leq\bar{\rho}\leq3.55$, $0.17\leq\bar{\rho}\leq2.32$, and $0.24\leq\bar{\rho}\leq2.10$. Additional simulations varying $a$ span $0.2\leq a\leq2.0$, $0.8\leq a\leq2.0$, and $1.0\leq a\leq2.0$ for $\nu=1,-1,-3$, respectively. (c) Estimated exponents $\gamma$ compared with Equation~\eqref{eq:dimensiondependentdiffusivity}. Theoretical results are shown for $d=1,2,3$ and $d=n$ as $n\to\infty$, together with simulation-based estimates for $d=1,2$. \cb
}
\label{fig:F3}
\end{figure}

Discrete simulations are performed by specifying $L$, $k$, $\eta$, $a$, $N$, $\nu$ and $\boldsymbol{x}(0)$.  Time-dependent solutions, either numerical or late-time exact solutions, can be presented by plotting the discrete density $\rho_n(t) = 1/\ell_n(t) = 1/(x_{n+1}(t) - x_{n}(t))$ as a function of the mid-point $(x_{n}(t) + x_{n+1}(t))/2$ for $n=0,1,\ldots,N-1$.  Results in Figure \ref{fig:F2} show  $\rho_n(t)$ as a function of time and position for $\nu = 1, -1, 3$.  In each case the initial step function in density relaxes to $\lim_{t \to \infty}\rho_n(t) = 1$, with the equally-spaced steady state $\bar{x}_n = nL/N$ for $n=0,1,\ldots,N$. \cbl These simulations consider the fundamental case $a=\bar{\ell}$ so that the resting spring length coincides with the long-time mean spring length and the steady state has zero prestress. Later results (Figure~\ref{fig:F3}) consider the same problem while varying $a$, so that the steady configuration is prestressed.  In these further cases $a<\bar{\ell}$ corresponds to extension whereas $a>\bar{\ell}$ corresponds to compression. The continuum--discrete agreement remains very good across these additional cases confirming that the PDE approximation is not restricted to zero-prestress configurations.   \cb \\

\noindent
\textit{Generalizing to higher dimensions}. 

In higher dimensions, cells are modeled by a tissue triangulation (Figure~\ref{fig:F1}c). Nodes interact through a network of links, giving
\begin{equation}\label{eq:discrete-higherdim-ode}
    \eta\, {\boldsymbol{r}_n}'(t) = -\sum_{k\sim n}f(\boldsymbol{r}_n-\boldsymbol{r}_k)\frac{\boldsymbol{r}_n-\boldsymbol{r}_k}{\|\boldsymbol{r}_n-\boldsymbol{r}_k\|},
\end{equation}
where $\boldsymbol{r}_n$ are node positions and $k\sim n$ sums over nodes $k$ adjacent to $n$. Restricting motion at domain boundaries (no flux) and linearizing about the steady state gives 
\begin{equation}\label{eq:discrete-higherdim}
    \eta \boldsymbol{\delta x}'(t) = - (L+B)\boldsymbol{\delta x},
\end{equation} 
where $\boldsymbol{\delta x}(t)$ collects all positional degrees of freedom, $L$ is the Laplacian matrix of the network weighted by the Jacobians of pairwise forces, and $B$ is a weighted degree matrix of boundary-restricted coordinates (see S.I.)~\cite{newman2018,mohar1997}. The solution of Equation~\eqref{eq:discrete-higherdim} sums terms that decrease to zero with rates $\exp\big[-(\tilde{\lambda}_p/\eta)t\big]$ set by the positive eigenvalues $\tilde{\lambda}_p$ of $L+B$, confirming late time relaxation is exponential. In 1D, $\tilde{\lambda}_p=\kappa\lambda_p$.

Microscopic conservation, a locality assumption, and dimensional arguments lead us to propose that the discrete model with Equation \eqref{eq:forcelaws} is related in any dimension to macroscopic nonlinear diffusion transport with nonlinear diffusion function~\cite{Barenblatt1996}
\begin{equation}\label{eq:GeneralDiffusivity}
D(\rho) = \dfrac{a^{\beta}\rho^{\gamma}}{\tau},
\end{equation}
where $\tau$ is a characteristic time scale (e.g. $\tau=\eta/k$).  Factors $a^\beta$ and $\rho^{\gamma}$ are included since $a$ and $\rho$ are related to two local length scales present in the discrete simulation.  While we could write Equation \eqref{eq:GeneralDiffusivity} as $D(\rho,a)$, we keep the $D(\rho)$ notation to be consistent with established literature~\cite{Barenblatt1996,OkuboLevin2001,McCue2019,Simpson2024}.  We propose Equation \eqref{eq:GeneralDiffusivity} under the standard assumption that no macroscopic length-scale contributes to $D(\rho)$, such as length-scales associated with the domain or initial condition. Simulation data confirms this.   Working in 1D, simulations involve a collection of cells with length $\ell_n > 0$ and density $\rho_n = 1/\ell_n$ with dimensions $[\rho] = \textrm{L}^{-1}$.  Simulations in 2D involve a collection of cells with triangular area $A_n > 0$ and density $\rho_n = 1/A_n$ with dimensions $[\rho] = \textrm{L}^{-2}$.  Since in $d$ dimensions, $[\rho] = \textrm{L}^{-d}$ and $[D(\rho)] = \textrm{L}^2\textrm{T}^{-1}$, we have
\begin{equation}\label{eq:dimensiondependentdiffusivity}
\gamma = \dfrac{\beta-2}{d},\qquad d=1,2,3,\ldots
\end{equation}
This scaling property suggests that implementing the same microscopic force model in different dimensions can lead to different macroscopic transport phenomena. For fixed microscopic scaling exponent $\beta$, the magnitude of the macroscopic density exponent $\gamma$ decreases with dimension $d$. One way of interpreting this is that linear diffusion $(\gamma = 0)$ is increasingly relevant in higher dimensions, $d \to \infty$. \cb  We now explore the relationship~\eqref{eq:dimensiondependentdiffusivity} using a combination of analysis and simulations.

\noindent
\textit{Macroscopic models.}

We seek a PDE description for the continuous macroscopic density $\rho(\mathbf{r},t)>0$ which we write in the form of a general nonlinear diffusion equation.
\begin{equation}\label{eq:3DPDE} 
\partial \rho /\partial t = \nabla \cdot \left( D(\rho)\nabla \rho\right).
\end{equation} 
\cbl We consider a nonlinear diffusion equation, commonly used to model population phenomena such as population pressure~\cite{Falco2023,Carrillo2019} and heterogeneous populations~\cite{Haridas2017,Carrillo2025}. \cb  We consider Equation \eqref{eq:3DPDE} with zero flux boundary conditions and general scalar diffusivity $D(\rho) \ge 0$.  Closed-form solutions of Equation \eqref{eq:3DPDE} are generally unavailable on a finite domain. The long-time steady solution is constant, $\lim_{t \to \infty} \rho(\mathbf{r},t)=  \bar{\rho}$ ( $\bar{\rho}=N/L$ in 1D), and late-time solutions are obtained by writing $ \rho(\mathbf{r},t) = \bar{\rho}+\varepsilon \rho_1(\mathbf{r},t)+\mathcal{O}(\varepsilon^2)$, where $0<\varepsilon\ll 1$. Assuming $D'(\rho)$ and $\rho_1(\mathbf{r},t)$ are $\mathcal{O}(1)$,  $\rho_1(\mathbf{r},t)$ undergoes linear diffusion 
\begin{equation}\label{eq:3DPMEApprox} 
\partial \rho_1/\partial t = D(\bar{\rho})\nabla^2 \rho_1, \end{equation}
where $D(\bar{\rho}) > 0$ is a constant that is given by the nonlinear diffusivity function evaluated at the long-time density.  

Solving Equation \eqref{eq:3DPMEApprox} in 1D on $0 < x < L$ with no-flux boundary conditions gives a leading eigenvalue approximation as $t \to \infty$ 
\begin{equation}\label{eq:DiffusinLateTime}
\rho_1(x,t) \propto \cos\left( \pi x/L\right)
\exp\left(
-D(\bar{\rho})\left(\pi/L\right)^2t
\right).
\end{equation}
Matching decay rates in Equations \eqref{eq:DiscreteLateTime} and \eqref{eq:DiffusinLateTime} gives
\begin{equation}\label{eq:1DDiffusivity}
D(\rho) = (k/\eta)a^{(1-\nu)}\rho^{-(1 + \nu)},
\end{equation}
which  is consistent with Equation \eqref{eq:GeneralDiffusivity}.  Since $D(\rho) \propto \rho^{- (1+\nu)}$, $D(\rho)$ decreases for $\nu > - 1$, increases for $\nu < - 1$, and is constant for $\nu =  - 1$, giving linear diffusion.   Equation \eqref{eq:1DDiffusivity} has $[D(\rho)] = \textrm{L}^2\textrm{T}^{-1}$ for all $\nu$ since $[\rho] = \textrm{L}^{-1}$ in 1D. Results in Figure \ref{fig:F2} compare discrete simulation data with numerical solutions of Equation \eqref{eq:3DPDE} with $D(\rho)=\rho^{-2}, \rho^0,\rho^2$ for $\nu = 1, - 1,  - 3$, respectively.  The close discrete--continuum agreement in Figure~\ref{fig:F2} confirms that matching late-time relaxation rates identifies $D(\rho)$, and the same late-time argument applies in higher dimensions. Solving Equation~\eqref{eq:3DPMEApprox} in 2D/3D using no-flux eigenfunctions of $-\nabla^2$ with eigenvalues $\Lambda_q>0$ gives exponential rates $\exp\big(\!-\!D(\bar{\rho})\Lambda_q t\big)$. On a rectangle, the rates are $-D(\bar{\rho})[(m\pi/L_x)^2+(n\pi/L_y)^2]$, with $m,n=0,1,2,\ldots$, and analogous expressions hold in 3D. Thus, for higher-dimensional relaxation dominated by the horizontal mode of the initial condition, the decay rate is $-D(\bar{\rho})(\pi/L_x)^2$, enabling the same relaxation-rate matching procedure in higher dimensions. We therefore obtain a general method to coarse-grain microscopic dynamics into nonlinear diffusion functions $D(\rho)$ based on comparing relaxation rates from the discrete and diffusion models with different steady-state densities. This method does not require continuum limits, which is particularly suited to finite systems.\\

\noindent
\cbl \textit{Revealing $D(\rho)$ using 1D simulation data.} 

To complement the theoretical results above, we develop a computational framework that provides a foundation for extending from 1D to higher dimensions, beginning in 1D where the relation between $f(\ell)$ and $D(\rho)$ is known from Equation~\eqref{eq:1DDiffusivity}. Figure~\ref{fig:F2}a considers $f(\ell)=k(\ell-a)$ with $N=100$ internal boundaries and $k=a=\eta=1$. From numerical solutions of the discrete model, we monitor $\delta x_n(t)=x_n(t)-\bar{x}_n$ for $n=21,51,91$. Once the leading mode dominates, $\delta x_n(t)\propto \exp(-\lambda_1 t)$, so $\log|\delta x_n(t)|$ is linear in $t$. Linear regression gives $\lambda_1=-9.8694\times10^{-4}$ in all three cases, matching $-(\pi/N)^2\approx-9.8696\times10^{-4}$. Using Equation~\eqref{eq:DiffusinLateTime}, this relaxation rate gives $D(1)=0.99998$ for $\bar{\rho}=1$, consistent with the known result $D(\rho)=\rho^{-2}$ and hence $D(1)=1$ for $k=\eta=1$~\cite{Murray2009,Murphy2019}. Repeating this procedure while varying $N$ and $\bar{\rho}=N/L$ reveals the density dependence of $D(\rho)$. For example, $N=200$ and $\bar{\rho}=2$ gives $D(2)=0.24996$, consistent with $D(2)=1/4$. Figure~\ref{fig:F3}a summarizes the resulting scalings: $\nu=1$ gives $D(\rho)\propto\rho^{-2.01}$, $\nu=-1$ gives $D(\rho)\propto\rho^{0.00}$, and $\nu=-3$ gives $D(\rho)\propto\rho^{2.00}$, all consistent with Equation~\eqref{eq:1DDiffusivity}. Additional discrete and continuum simulations (not shown), using different domains, initial conditions, and choices of $\nu,a,k,L,\eta$, show similarly close agreement between the discrete model and Equation~\eqref{eq:3DPDE} with diffusivity given by Equation~\eqref{eq:1DDiffusivity}. Together, these results support our proposition that the local length scales $a$ and $\rho$ determine the diffusivity, Equation~\eqref{eq:GeneralDiffusivity}. \cb \\

\noindent
\textit{Revealing $D(\rho)$ using 2D simulation data.} 

We perform 2D simulations on a rectangle, $0<x<L_x$, $0<y<L_y$.  Visualization of the 2D simulations are provided on GitHub~\cite{Simpson2026GitHub}. Initial cell-boundary locations are sampled from a prescribed density profile: lower density for $x<0.4L_x$, higher density for $x>0.6L_x$, and a smooth transition for $0.4L_x\leq x\leq 0.6L_x$. The 2D network is defined by the Delaunay triangulation of the initial node locations. Cell density is $\rho_n(t)=1/A_n(t)$, where $A_n(t)$ is the area of the $n$th triangle (Figure~\ref{fig:F1}c).   While nodes are free to move in any direction (except at domain boundaries for no-flux conditions), the initial cell density depends upon the horizontal coordinate alone. Simulations are performed by solving Equation~\eqref{eq:discrete-higherdim-ode} numerically without remeshing, using $\eta=1$.  All simulations involve a  fixed number of triangular cells, $Q$. The system relaxes to a uniform steady state with density $\bar{\rho}=Q/(L_x L_y)$.  We estimate the relaxation rate by randomly selecting 20 cell nodes and track the horizontal position of the $i$th node, $x_i(t)$, defining $\delta x_i(t)=x_i(t)-x_i(T)$, where $T$ is a sufficiently late simulation time.  Once the leading mode dominates, $\log |\delta x_i(t)|$ is linear in $t$ with slope $\tilde\lambda_1$. Averaging these slopes from each sampled node and verifying small sample variance, we obtain an estimate of $\tilde\lambda_1$, which we match with $-D(\bar{\rho})(\pi/L_x)^2$. Figure~\ref{fig:F3}b shows that $D(\rho)\propto \rho^\gamma$ for $\nu=1,-1,-3$, all with $a=1$. The exponent $\gamma$ depends on dimension even when $f(\ell)$ and $\eta$ are fixed. For example, $\nu=1$ gives $D(\rho)\propto \rho^{-2.01}$ in 1D and $D(\rho)\propto \rho^{-1.11}$ in 2D, whereas $\nu=-3$ gives $D(\rho)\propto \rho^{2.00}$ in 1D and $D(\rho)\propto \rho^{0.47}$ in 2D, confirming $|\gamma|$ decreases with $d$.

\cbl The mechanical response of disordered spring and fibre networks depends strongly on their coordination number and state of prestress~\cite{Picu2011,Broedersz2014}. For networks with central-force interactions, Maxwell counting identifies the isostatic threshold as $z_{\rm c}=2d$, so that networks below this threshold are generically floppy unless stabilized by additional mechanisms, whereas networks above it are rigid~\cite{Broedersz2011,Arzash2019,Merkel2019,Cui2019,Damavandi2022,Damavandi2022b}. Our 2D  Delaunay networks have a mean coordination number approaching $z=6$ in the bulk, placing them above the central-force isostatic threshold $z_{\rm c}=4$.  Consequently, we do not consider a floppy-to-rigid transition here. Nevertheless, individual links can remain prestressed at mechanical equilibrium, and this prestress contributes to the linearized network dynamics as we will now illustrate. \cb

To generalize results in Figure~\ref{fig:F3}a--b obtained with constant $a=1$, we vary both $a$ and $\nu$ in a similar suite of 1D/2D simulations. Figure~\ref{fig:F3}c show that exponents $\beta$ and $\gamma$ are related by the proposed universal relationship, Equation~\eqref{eq:GeneralDiffusivity}.  In general the degree of nonlinearity in the coarse-grained $D(\rho)$ decreases with dimension since $|\gamma|$ decreases in moving from 1D to 2D. Equation~\eqref{eq:GeneralDiffusivity} has $\gamma \to 0$ as $d \to \infty$, meaning that linear diffusion becomes increasingly relevant in higher dimensions.

In summary, we propose a novel method to coarse-grain microscopic dynamics into macroscopic diffusion that does not rely on continuum limits. Our results show that dimensionality is not simply a geometric detail: it fundamentally changes the macroscopic law obtained after coarse-graining. This principle is familiar from random-walk theory, where return probabilities and recurrence depend upon dimension~\cite{Hughes1995,Redner2001}, and from renormalization-group methods, where changing dimension can alter effective large-scale descriptions~\cite{Goldenfeld1990,Bricmont1991}. We reveal an analogous phenomenon in tissue mechanics. The same microscopic interaction law does not generate a dimension-independent continuum limit; instead, dimension reshapes the effective nonlinear diffusivity, changing the exponent in $D(\rho)\propto\rho^\gamma$.  Importantly, insights from 1D  models cannot be simply assumed to hold unchanged in higher dimensions, and outcomes from linear models in 1D (Hooke's law) masks critical details about how microscopic length scales map to emergent descriptions of tissue-scale transport and relaxation phenomena.  

\cbl Our results also have implications for experiments performed under different geometrical constraints. Epithelial tissues confined to narrow adhesive tracks or channels provide effectively 1D systems, and experiments show that geometrical confinement can strongly influence collective migration and force transmission~\cite{Vedula2012}. Conventional epithelial monolayers cultured on planar substrates provide an effectively 2D setting in which cellular tractions, intercellular stresses and tissue relaxation can be quantified directly~\cite{Trepat2009,SerraPicamal2012}. In 3D, tumour spheroids and organoids provide complementary experimental systems for studying tissue growth and mechanics within multicellular aggregates~\cite{Murphy2022,PerezGonzalez2021}. Our results suggest that continuum descriptions inferred from these different experimental geometries need not be directly interchangeable.  Even the same microscopic cell--cell force law, the density dependence of the effective diffusivity changes with dimensionality. Thus, comparing relaxation dynamics across confined channels, planar monolayers and 3D aggregates could provide a direct experimental route for testing the dimension-dependent coarse-graining predicted here. \cb

\cbl One important extension of this work would be to systematically vary network connectivity and prestress, including sub-isostatic networks in which floppy modes can be stabilized by bending or prestress and rigidity can emerge through stress-induced or geometric mechanisms~\cite{Broedersz2011,Arzash2019,Merkel2019,Cui2019,Damavandi2022,Damavandi2022b}. This would allow us to examine how proximity to a rigidity transition influences the relationship between microscopic force laws, macroscopic transport, and late-time relaxation. \cb\\

\noindent 
\textit{Data availability:} Open source software and all data required to generate results are available on GitHub~\cite{Simpson2026GitHub}.\\

\noindent 
\textit{Acknowledgments:} Supported by ARC DP230100025. {\cbl We thank two referees for helpful comments.}

\bibliography{references} 

@article{Arzash2019,
  author  = {Arzash, S. and Shivers, J. L. and Licup, A. J. and Sharma, A. and MacKintosh, F. C.},
  title   = {Stress-stabilized subisostatic fiber networks in a ropelike limit},
  journal = {Phys. Rev. E},
  volume  = {99},
  pages   = {042412},
  year    = {2019},
  doi     = {10.1103/PhysRevE.99.042412}
}

@book{Barenblatt1996,
  author = {Barenblatt, G. I.},
  title = {Scaling, Self-similarity, and Intermediate Asymptotics},
  year = {1996},
  address = {Cambridge},
  publisher = {CUP},
  doi = {10.1017/CBO9781107050242}
}

@article{Broedersz2011,
  author  = {Broedersz, C. P. and Mao, X. and Lubensky, T. C. and MacKintosh, F. C.},
  title   = {Criticality and isostaticity in fibre networks},
  journal = {Nat. Phys.},
  volume  = {7},
  pages   = {983--988},
  year    = {2011},
  doi     = {10.1038/nphys2127}
}

@article{Broedersz2014,
  author  = {Broedersz, C. P. and MacKintosh, F. C.},
  title   = {Modeling semiflexible polymer networks},
  journal = {Rev. Mod. Phys.},
  volume  = {86},
  pages   = {995--1036},
  year    = {2014},
  doi     = {10.1103/RevModPhys.86.995}
}

@book{Bear1972,
  author = {Bear, Jacob},
  title = {Dynamics of Fluids in Porous Media},
  year = {1972},
  address = {New York},
  publisher = {American Elsevier}
}

@article{Buenzli2025,
  title = {Mechanical cell interactions on curved interfaces},
  author = {Buenzli, Pascal R. and Kuba, Shahak and Murphy, Ryan J. and Simpson, Matthew J.},
  journal = {Bull. Math. Biol.},
  volume = {87},
    pages = {28},
  year = {2025},
  doi = {10.1007/s11538-024-01406-w},
}

@article{Carrillo2019,
  author  = {Carrillo, Jos{\'e} A. and Murakawa, Hideki and Sato, Makoto and Togashi, Hideru and Trush, Olena},
  title   = {A population dynamics model of cell-cell adhesion incorporating population pressure and density saturation},
  journal = {J. Theor. Biol.},
  volume  = {474},
  pages   = {14--24},
  year    = {2019},
  doi     = {10.1016/j.jtbi.2019.04.023}
}

@article{Carrillo2025,
  author  = {Carrillo, Jos{\'e} A. and Lorenzi, Tommaso and Macfarlane, Fiona R.},
  title   = {Spatial segregation across travelling fronts in individual-based and continuum models for the growth of heterogeneous cell populations},
  journal = {Bull. Math. Biol.},
  volume  = {87},
  pages   = {77},
  year    = {2025},
  doi     = {10.1007/s11538-025-01452-y}
}

@article{Cui2019,
  author  = {Cui, B. and Ruocco, G. and Zaccone, A.},
  title   = {Theory of elastic constants of athermal amorphous solids with internal stresses},
  journal = {Granul. Matter},
  volume  = {21},
  pages   = {69},
  year    = {2019},
  doi     = {10.1007/s10035-019-0916-4}
}

@article{Damavandi2022,
  author  = {Damavandi, Ojan Khatib and Hagh, Varda F. and Santangelo, Christian D. and Manning, M. Lisa},
  title   = {Energetic rigidity. I. A unifying theory of mechanical stability},
  journal = {Phys. Rev. E},
  volume  = {105},
  pages   = {025003},
  year    = {2022},
  doi     = {10.1103/PhysRevE.105.025003}
}

@article{Damavandi2022b,
  author  = {Damavandi, Ojan Khatib and Hagh, Varda F. and Santangelo, Christian D. and Manning, M. Lisa},
  title   = {Energetic rigidity. II. Applications in examples of biological and underconstrained materials},
  journal = {Phys. Rev. E},
  volume  = {105},
  pages   = {025004},
  year    = {2022},
  doi     = {10.1103/PhysRevE.105.025004}
}

@article{Falco2023,
  author  = {Falc{\'o}, Carles and Cohen, Daniel J. and Carrillo, Jos{\'e} A. and Baker, Ruth E.},
  title   = {Quantifying tissue growth, shape and collision via continuum models and Bayesian inference},
  journal = {J. R. Soc. Interface},
  volume  = {20},
  number  = {204},
  pages   = {20230184},
  year    = {2023},
  doi     = {10.1098/rsif.2023.0184}
}

@article{Fozard2009,
author  = {Fozard, J. A. and Byrne, H. M. and Jensen, O. E. and King, J. R.},
title   = {Continuum approximations of individual-based models for epithelial monolayers},
journal = {Math. Med. Biol.},
volume  = {27},
number  = {1},
pages   = {39--74},
year    = {2009},
doi     = {10.1093/imammb/dqp015}
}

@book{Haberman2013,
  author    = {R. Haberman},
  title     = {Applied Partial Differential Equations with Fourier Series and Boundary Value Problems},
  publisher = {Pearson},
  year      = {2013}
}

@article{Haridas2017,
  author  = {Haridas, Parvathi and Penington, Catherine J. and McGovern, Jacqui A. and McElwain, D. L. Sean and Simpson, Matthew J.},
  title   = {Quantifying rates of cell migration and cell proliferation in co-culture barrier assays reveals how skin and melanoma cells interact during melanoma spreading and invasion},
  journal = {J. Theor. Biol.},
  volume  = {423},
  pages   = {13--25},
  year    = {2017},
  doi     = {10.1016/j.jtbi.2017.04.017}
}

@article{Kouachi2006,
  author  = {Kouachi, Said},
  title   = {Eigenvalues and eigenvectors of tridiagonal matrices},
  journal = {Electron. J. Linear Algebra},
  volume  = {15},
  pages   = {115--133},
  year    = {2006},
  doi     = {10.13001/1081-3810.1223},
  url     = {https://eudml.org/doc/126465}
}

@article{Merkel2019,
  author  = {Merkel, M. and Baumgarten, K. and Tighe, B. P. and Manning, M. L.},
  title   = {A minimal-length approach unifies rigidity in underconstrained materials},
  journal = {Proc. Natl. Acad. Sci. U.S.A.},
  volume  = {116},
  pages   = {6560--6568},
  year    = {2019},
  doi     = {10.1073/pnas.1815436116}
}

@article{McCue2019,
  author = {McCue, Scott W. and Jin, Wang and Moroney, Timothy J. and Lo, Kai-Yin and Chou, Shih-En and Simpson, Matthew J.},
  title = {Hole-closing model reveals exponents for nonlinear degenerate diffusivity functions in cell biology},
  journal = {Physica D},
  volume = {398},
  pages = {130--140},
  year = {2019},
  doi = {10.1016/j.physd.2019.06.005}
}

@article{Mirams2013,
author  = {Mirams, Gary R. and Arthurs, Christopher J. and Bernabeu, Miguel O. and Bordas, Rafel and Cooper, Jonathan and Corrias, Alberto and Davit, Yohan and Dunn, Sara-Jane and Fletcher, Alexander G. and Harvey, Daniel G. and Marsh, Michael E. and Osborne, James M. and Pathmanathan, Pras and Pitt-Francis, Joe and Southern, James and Zemzemi, Nejib and Gavaghan, David J.},
  title   = {Chaste: An open source {C++} library for computational physiology and biology},
  journal = {PLoS Comput. Biol.},
  year    = {2013},
  volume  = {9},
  number  = {3},
  pages   = {e1002970},
  doi     = {10.1371/journal.pcbi.1002970}
}

@article{Murphy2019,
    author = {Murphy, R. J. and Buenzli, P. R. and Baker, R. E. and Simpson, M. J.},
    title = "{A one-dimensional individual-based mechanical model of cell movement in heterogeneous tissues and its coarse-grained approximation}",
    journal = {Proc. R. Soc. A},
    volume = {475},
    pages = {20180838},
    year = {2019},
    doi = {10.1098/rspa.2018.0838},
}

@article{Murphy2020,
    author = {Murphy, R. J. and Buenzli, P. R. and Baker, R. E. and Simpson, M. J.},
    title = "{Mechanical cell competition in heterogeneous epithelial tissues}",
    journal = {Bull. Math. Biol.},
    volume = {475},
    pages = {130},
    year = {2019},
    doi = {10.1007/s11538-020-00807-x},
}

@article{Murphy2022,
  author  = {Murphy, Ryan J. and Browning, Alexander P. and Gunasingh, Gency
             and Haass, Nikolas K. and Simpson, Matthew J.},
  title   = {Designing and interpreting {4D} tumour spheroid experiments},
  journal = {Commun. Biol.},
  volume  = {5},
  pages   = {91},
  year    = {2022},
  doi     = {10.1038/s42003-022-03018-3}
}

@book{Murray2002,
  author = {Murray, J. D.},
  title = {Mathematical Biology I: An Introduction},
  volume = {17},
  edition = {3},
  year = {2002},
  address = {New York},
  publisher = {Springer},
  doi = {10.1007/b98868}
}

@article{Murray2009,
  title = {From a discrete to a continuum model of cell dynamics in one dimension},
  author = {Murray, P. J. and Edwards, C. M. and Tindall, M. J. and Maini, P. K.},
  journal = {Phys. Rev. E},
  volume = {80},
  pages = {031912},
  numpages = {10},
  year = {2009},
  doi = {10.1103/PhysRevE.80.031912},
}

@article{Murray2011,
author  = {Philip J Murray and Alex Walter and Alexander G Fletcher and Carina M Edwards and Marcus J Tindall and Philip K Maini},
title   = {Comparing a discrete and continuum model of the intestinal crypt},
journal = {Phys. Biol.},
year    = {2011},
volume  = {8},
pages   = {026011},
doi     = {10.1088/1478-3975/8/2/026011}
}

@article{Murray2012,
  title = {Classifying general nonlinear force laws in cell-based models via the continuum limit},
  author = {Murray, P. J. and Edwards, C. M. and Tindall, M. J. and Maini, P. K.},
  journal = {Phys. Rev. E},
  volume = {85},
  pages = {021921},
  year = {2012},
  doi = {10.1103/PhysRevE.85.021921},
}

@article{Odell1981,
  author  = {Odell, G. M. and Oster, G. and Alberch, P. and Burnside, B.},
  title   = {The mechanical basis of morphogenesis. {I}. Epithelial folding and invagination},
  journal = {Dev. Biol.},
  year    = {1981},
  volume  = {85},
  number  = {2},
  pages   = {446--462},
  doi     = {10.1016/0012-1606(81)90276-1}
}

@book{OkuboLevin2001,
author = {Okubo, A. and Levin, S. A.},
title = {Diffusion and Ecological Problems: Modern Perspectives},
year = {2001},
address = {New York},
publisher = {Springer},
doi = {10.1007/978-1-4757-4978-6}
}

@article{Oster1983,
  author  = {Oster, G. F. and Murray, J. D. and Harris, A. K.},
  title   = {Mechanical aspects of mesenchymal morphogenesis},
  journal = {J. Embryol. Exp. Morphol.},
  year    = {1983},
  volume  = {78},
  pages   = {83--125}
}

@article{PerezGonzalez2021,
  author  = {P{\'e}rez-Gonz{\'a}lez, Carlos and Ceada, Gerardo and Greco, Francesco
             and Matej{\v{c}}i{\'c}, Marija and G{\'o}mez-Gonz{\'a}lez, Manuel
             and Castro, Natalia and Menendez, Anghara and Kale, Sohan
             and Krndija, Denis and Clark, Andrew G. and Gannavarapu, Venkata Ram
             and {\'A}lvarez-Varela, Adri{\'a}n and Roca-Cusachs, Pere
             and Batlle, Eduard and Matic Vignjevic, Danijela
             and Arroyo, Marino and Trepat, Xavier},
  title   = {Mechanical compartmentalization of the intestinal organoid enables crypt folding and collective cell migration},
  journal = {Nat. Cell Biol.},
  volume  = {23},
  pages   = {745--757},
  year    = {2021},
  doi     = {10.1038/s41556-021-00699-6}
}

@article{PittFrancis2009,
  author  = {Pitt-Francis, Joe and Pathmanathan, Pras and Bernabeu, Miguel O. and Bordas, Rafel and Cooper, Jonathan and Fletcher, Alexander G. and Mirams, Gary R. and Murray, Peter and Osborne, James M. and Walter, Alex and Chapman, S. Jonathan and Garny, Alan and van Leeuwen, Ingeborg M. M. and Maini, Philip K. and Rodr{\'i}guez, Blanca and Waters, Sarah L. and Whiteley, Jonathan P. and Byrne, Helen M. and Gavaghan, David J.},
  title   = {Chaste: A test-driven approach to software development for biological modelling},
  journal = {Comput. Phys. Commun.},
  year    = {2009},
  volume  = {180},
  number  = {12},
  pages   = {2452--2471},
  doi     = {10.1016/j.cpc.2009.07.019}
}

@article{SerraPicamal2012,
  author  = {Serra-Picamal, Xavier and Conte, Vito and Vincent, Romaric
             and Anon, Ester and Tambe, Dhananjay T. and Bazellieres, Elsa
             and Butler, James P. and Fredberg, Jeffrey J. and Trepat, Xavier},
  title   = {Mechanical waves during tissue expansion},
  journal = {Nat. Phys.},
  volume  = {8},
  pages   = {628--634},
  year    = {2012},
  doi     = {10.1038/nphys2355}
}

@article{Simpson2015,
  author  = {Simpson, Matthew J. and Baker, Ruth E.},
  title   = {Exact calculations of survival probability for diffusion on growing lines, disks, and spheres: The role of dimension},
  journal = {J. Chem. Phys.},
  volume  = {143},
  number  = {9},
  pages   = {094109},
  year    = {2015},
  doi     = {10.1063/1.4929993}
}

@article{Simpson2024,
  author = {Simpson, Matthew J. and McCue, Scott W.},
  title = {{Fisher--KPP}-type models of biological invasion: open source computational tools, key concepts and analysis},
journal = {Proc. R. Soc. A},
  volume = {480},
  number = {2294},
  pages = {20240186},
  year = {2024},
  doi = {10.1098/rspa.2024.0186}
}

@article{Simpson2025,
  author  = {Simpson, Matthew J. and Plank, Michael J.},
  title   = {Inference and prediction for stochastic models of biological populations undergoing migration and proliferation},
  journal = {J. R. Soc. Interface},
  year    = {2025},
  volume  = {22},
  number  = {231},
  pages   = {20250536},
  doi     = {10.1098/rsif.2025.0536}
}

@misc{Simpson2026GitHub,
  author       = {Simpson, Matthew J},
title        = {{Code and data}},
  year         = {2026},
  publisher    = {GitHub},
  howpublished = {\url{https://github.com/ProfMJSimpson/tissue}}
}

@article{Trepat2009,
  author  = {Trepat, Xavier and Wasserman, Michael R. and Angelini, Thomas E.
             and Millet, Emil and Weitz, David A. and Butler, James P.
             and Fredberg, Jeffrey J.},
  title   = {Physical forces during collective cell migration},
  journal = {Nat. Phys.},
  volume  = {5},
  pages   = {426--430},
  year    = {2009},
  doi     = {10.1038/nphys1269}
}

@book{Vazquez2007,
  author = {V{\'a}zquez, Juan Luis},
  title = {The Porous Medium Equation: Mathematical Theory},
  year = {2007},
  address = {Oxford},
  publisher = {OUP},
  doi = {10.1093/acprof:oso/9780198569039.001.0001}
}

@book{Hughes1995,
  author    = {Hughes, Barry D.},
  title     = {Random Walks and Random Environments},
  publisher = {OUP},
  address   = {Oxford},
  year      = {1995}
}

@article{Goldenfeld1990,
  author  = {Goldenfeld, Nigel and Martin, Olivier and Oono, Y. and Liu, Fong},
  title   = {Anomalous dimensions and the renormalization group in a nonlinear diffusion process},
  journal = {Phys. Rev. Lett.},
  year    = {1990},
  volume  = {64},
  number  = {12},
  pages   = {1361--1364},
  doi     = {10.1103/PhysRevLett.64.1361}
}

@article{Bricmont1991,
  author  = {Bricmont, J. and Kupiainen, A.},
  title   = {Renormalization group for diffusion in a random medium},
  journal = {Phys. Rev. Lett.},
  year    = {1991},
  volume  = {66},
  number  = {13},
  pages   = {1689--1692},
  doi     = {10.1103/PhysRevLett.66.1689}
}

@book{newman2018,
author = {Newman, Mark},
title = {Networks: An Introduction},
year = {2018},
address = {Oxford},
publisher = {OUP},
doi = {10.1093/oso/9780198805090.001.0001}
}

@incollection{mohar1997,
  title     = {Some applications of Laplace eigenvalues of graphs},
  author    = {Mohar, Bojan},
  booktitle = {Graph Symmetry: Algebraic Methods and Applications},
  editor    = {Hahn, Geňa and Sabidussi, Gert},
  series    = {NATO ASI Series C: Mathematical and Physical Sciences},
  volume    = {497},
  pages     = {227--275},
  year      = {1997},
  publisher = {Springer},
  doi       = {10.1007/978-94-015-8937-6_6}
}

@article{Picu2011,
  author  = {Picu, R. C.},
  title   = {Mechanics of random fiber networks---a review},
  journal = {Soft Matter},
  volume  = {7},
  pages   = {6768--6785},
  year    = {2011},
  doi     = {10.1039/C1SM05022B}
}

@book{Redner2001,
  author    = {Redner, Sidney},
  title     = {A Guide to First-Passage Processes},
  publisher = {CUP},
  address   = {Cambridge},
  year      = {2001},
  doi       = {10.1017/CBO9780511606014}
}

@article{Vedula2012,
  author  = {Vedula, Sri Ram Krishna and Leong, Man Chun and Lai, Tan Lei
             and Hersen, Pascal and Kabla, Alexandre J. and Lim, Chwee Teck
             and Ladoux, Beno{\^i}t},
  title   = {Emerging modes of collective cell migration induced by geometrical constraints},
  journal = {Proc. Natl. Acad. Sci. U.S.A.},
  volume  = {109},
  number  = {32},
  pages   = {12974--12979},
  year    = {2012},
  doi     = {10.1073/pnas.1119313109}
}

\end{document}